\documentclass[seceq.ptp-prep]{ptptex}
\usepackage{graphicx}
\usepackage{amsmath}
\usepackage{txfonts}
\newcommand{\mbf}[1]{\boldsymbol{#1}}

\title{Q-Deformed Bi-Local Fields}

\author{
Shigefumi {\sc Naka}, Haruki {\sc Toyoda} and Aiko {\sc Kimishima}, 
}

\inst{
Department of Physics, College of Science and Technology \\
Nihon University, Tokyo 101-8308,Japan
}

\abst{We study the q-deformation of the bi-local system, two particle system, bounded by a relativistic harmonic oscillator type of potential from both points of view of mass spectra and the behavior of scattering amplitudes. In particular, we try to formulate the deformation so that $P^2$, the square of center of mass momenta, enters into the deformation parameters of relative coordinates. As a result, the wave equation of the bi-local system becomes non-linear one with respect to $P^2$; then, the propagator of the bi-local system suffers significant change so as to get a convergent self energy to second order. 
}

\begin{document}

\maketitle

\section{Introduction}

The aim of non-local field theories proposed originally by Yukawa\cite{non-local} was twofold: firstly, characteristic properties of elementary particles such as mass spectrum of hadrons are to be derived from their extended structure. Secondly, those fields are expected to save the divergence difficulty, which is inherent in the local field theories with local interactions. The bi-local field theory\cite{bi-local} was the first attempt by Yukawa in this line of thought. As for the bi-local fields, the first aim has been studied by many authors in the context of effective relativistic two-particle systems of quark and anti-quark bound systems. In particular, the two-particle systems bounded by a relativistic harmonic oscillator potential were desirable to get a linear mass square spectrum associated with the Regge behavior in their scattering amplitude\cite{Barger-Cline}\cite{Regge}.

On the other hand, the second aim was rather unsuccessful mainly because the bi-local fields are reduced to superposition of infinite local fields with different masses, although some attempts claimed that the second order self-energy becomes convergent associated with the direction of center of mass momenta. In addition to this, the problem of unitarily of scattering matrix and that of the causality are also serious problems for those fields, since the bi-local system allows time-like relative motions in general. Usually, such a degree of freedom is frozen by an additional subsidiary condition\cite{Takabayasi}, which is not always successful, however, for interacting cases. This situation may be different from that of string models which are characterized by the Virasoro condition associated with the parameterization invariance in such an extended model; nevertheless the study of bi-local field theories does not come to end, since a small change in models, sometimes, will cause a significant change in their physical properties.

Under these backgrounds, the purpose of this paper is to study the q-deformation\cite{Macfarlane}\cite{Wess}\cite{Sogami}\cite{Quantum-Groups} of a bi-local system characterized by the relativistic harmonic oscillator potential, since the q-deformation is well defined for harmonic oscillator systems. In a previous paper\cite{5-dimension}, we have studied a q-deformed 5-dimensional spacetime such that the extra dimension generates a harmonic oscillator type of potential for particles embedded in that spacetime. Then, the propagator of particles in that spacetime acquires significant convergent property by requiring that the 4-dimensional spacetime variables and the extra dimensional one are mixed by the deformation. It is, then, happen that the spacetime variables become non-commutative between 4-dimensional components and the fifth one. We can expect the same situation in the bi-local system if we carry out the deformation of the relative variables in the bi-local system on a parallel with the extra dimension in the q-deformed 5-dimensional spacetime.

In the next section, we try to formulate the bi-local system with q-deformed relative coordinates. In that case, we have to define the deformation carefully to keep the covariance, since the q-deformation of harmonic oscillator is ambiguous other than one dimensional space. In \S 3, the interaction of the bi-local field is discussed within the level of calculating Feynman diagrams. Some scattering amplitudes between the bi-local system and external scalar fields is also studied paying attention to their Regge behavior. We also calculate a self-energy type of diagram to study the convergence of the model to the second order. \S 4. is devoted to the summary and discussions; and in Appendix A, we also give the discussions on the representation of q-deformation in N-dimensional space.

\section{Formulation}

The classical action of equal mass two-particle system that leads to the standard set of bi-local field equations is simply given by
\begin{equation}
 S=\sum_{i=1}^2\frac{1}{2}\int d\tau\left\{\frac{\dot{x}^{(i)2}}{e_i}+e_i\left(m^2+V(\bar{x}) \right) \right\},  ~(\bar{x}=x^{(1)}-x^{(2)}) \label{BL action}
\end{equation}
where $e_i(\tau),(i=1,2)$ are gauge variables, einbein's, that assure the reparametrization invariance with respect to $\tau$. Namely, it is assumed that the $e_i$ transforms
\begin{equation}
 e_i(\tau)=\left(\frac{df(\tau')}{d\tau'}\right)^{-1}e_i(\tau'), \label{gauge transformation}
\end{equation}
according as $\tau=f(\tau')$. By definition, the momentum conjugate to $x^{(i)}$ is given by $p^{(i)}=\frac{\delta S}{\delta x^{(i)}}=\frac{1}{e_i}\dot{x}^{(i)}$, and so, the variation of $S$ with respect to $e_i$ gives rise to the constraints
\footnote{
If we eliminate $e_i$ by using these constraints, the action of the bi-local system can be rewritten as ${\displaystyle S=\sum_i\int d\tau \sqrt{(m^2+V(\bar{x}))\dot{x}^{(i)2}} }$  ~ provided that ~ $m^2+V(\bar{x})\neq 0$.
}
\begin{equation}
 \frac{\delta S}{\delta e_i}=-\left\{p^{(i)2}-\left( m^2+V(\bar{x}) \right) \right\} =0,~(i=1,2). \label{BLC}
\end{equation}
The sum of these constraints becomes
\begin{equation}
 \sum_{i=1}^2(p^{(i)2}-m^2)-2V(\bar{x})=\frac{1}{2}P^2+2(\bar{p}^2-m^2)-2V(\bar{x})=0, \label{BLC-1}
\end{equation}
where $P=p^{(1)}+p^{(2)}$ and $\bar{p}=\frac{1}{2}(p^{(1)}-p^{(2)})$. On the other hand, the subtraction between two constraints leads to the constraint
\begin{equation}
 P\cdot\bar{p}=0 , \label{BLC-2}
\end{equation}
which can eliminate the degree of freedom of the relative time $\bar{x}^0$ at the rest frame $P=(P^0, \mbf{0})$. 

In practice, however, compatibility between the constraints (\ref{BLC-1}) and (\ref{BLC-2}) dependents on the functional form of $V(\bar{x})$. A simple way to remove this problem is to make the substitutions $\bar{x}^\mu \rightarrow \bar{x}_\perp^\mu$ and $\bar{p}^\mu \rightarrow \bar{p}_\perp^\mu$, where $\bar{x}_\perp =(\eta_{\mu\nu}-\frac{P_\mu P_\nu}{P^2})\bar{x}^\nu$ and so on. As for the covariant harmonic oscillator potential $V(\bar{x})=-\kappa^2 \bar{x}^2$, there is another way, which is similar to the Gupta-Bleuler formalism in Q.E.D.. In this case, if we define the oscillator variables $(a^\dagger,a)$ in q-number theory by
\begin{equation}
 \bar{x}=\frac{1}{\sqrt{2\kappa}}(a^\dagger + a),~~~~
 \bar{p}=i\sqrt{\frac{\kappa}{2}}(a^\dagger - a) ,
\end{equation}
then Eq(\ref{BLC-1}) and Eq.(\ref{BLC-2}) can be understood as the wave equation for the bi-local system and its subsidiary condition such that
\begin{eqnarray}
  & \left(\alpha^\prime P^2 + \frac{1}{2}\{ a^\dagger_\mu,a^\mu \}-\omega \right)|\Phi \rangle =0, \label{wave-eq} & \\
 & P\cdot a|\Phi \rangle = 0, & \label{subsidiary}
\end{eqnarray}
where $\alpha^\prime=\frac{1}{8\kappa}$ and $\omega=\frac{m^2}{2\kappa}$. The compatibility between Eqs.(\ref{wave-eq}) and (\ref{subsidiary}) can be verified easily; and further, the Eq.(\ref{BLC-2}) holds in the sense of the expectation value $\langle\Phi|P\cdot\bar{p}|\Phi\rangle = 0$. 

Now, let us consider the q-deformation, the mapping $(a_\mu,a_\mu^\dagger)\rightarrow (A_\mu,A_\mu^\dagger)$, of this bi-local system. In our standpoint, the deformation is a way getting a new wave equation without changing the geometrical meaning of particle's cordinates $x^{(i)},^(i=1,2)$. As shown in Appendix A, however, the way of q-deformation is not unique; and we follow the case (iii) in Appendix A defined by the mapping
\begin{eqnarray}
 A_\mu=a_\mu\sqrt{\frac{[N]_q}{N_\mu}},~~A_\mu^\dagger=\sqrt{\frac{[N]_q}{N_\mu}}a_\mu^\dagger ~, \label{mapping} \\
 (N=-a_\mu^\dagger a^\mu;~N_0=-a_0^\dagger a_0,N_i=a_i^\dagger a_i). \nonumber
\end{eqnarray}
Here $[N]_q$ is the q-deformed number operator defined in Eq.(\ref{[N]}) with
\begin{equation}
 \alpha=1~~~{\rm and}~~~\beta=\beta_0-\alpha_0^\prime P^2~~,(\alpha_0^\prime >0). \label{beta}
\end{equation}
We note that the $\beta$ in Eq.(\ref{[N]}) may be a q-number variable commuting with oscillator variables. In the bi-local system, Eq.(\ref{beta}) gives a possible form of q-number $\beta$, which does not break the Lorentz covariance and the transnational invariance of the resultant wave equation. Then, the q-deformed counterpart of Eq.(\ref{wave-eq}) becomes
\begin{equation}
 \left(\alpha^\prime P^2 + \frac{1}{2}\{ A^\dagger_\mu,A^\mu \}(P^2,N)-\omega \right)|\Phi \rangle = 0, \label{q-wave-eq} 
\end{equation}
where
\begin{equation}
 \frac{1}{2}\{ A^\dagger_\mu,A^\mu \}(P^2,N)=-2\frac{\sinh\{(N+\beta_0-\alpha_0^\prime P^2 +\frac{1}{2})\log q\}}{\sinh(\frac{1}{2}\log q)}  \label{q-mass}
\end{equation}
The subsidiary condition (\ref{subsidiary}) is again compatible with the q-deformed wave equation (\ref{q-wave-eq}). Further, if we take the limit $q\rightarrow 1$, then Eq.(\ref{q-wave-eq}) will be reduced to Eq.(\ref{wave-eq}) provided that $\alpha_0^\prime=\frac{3}{4}\alpha^\prime$ and $\beta_0=\frac{3}{4}\omega-4$. It should also be noted that the resultant wave equation is covariant under the Lorentz transformation though $(A_\mu,A_\mu^\dagger)$ do not transform as four vectors.

Now, the mass square of the q-deformed bi-local system is determined by solving $m_n^2=-\frac{1}{2\alpha^\prime}\{A,A^\dagger\}(m_n^2)+\frac{\omega}{\alpha^\prime}$ for each $n(=0,1,2,\cdots)$, the eigenvalue of $N$. The $m_n^2$ can be solved uniquely under the sign of $\alpha_0^\prime$ in Eq.(\ref{beta}) so that there exist no spacelike solutions as in Fig.1. It should also be noted that the operator acting on the physical state $|\Phi \rangle$ in Eq.(\ref{q-wave-eq}) behaves as $(\cdots)=\sum_{k\geq 1}c_k(n)(P^2-m_n^2)^k$ with $c_1(n)=\frac{\partial}{\partial P^2}(\cdots)|_{P^2=m_n^2}\neq 0$ near the on mass-shell points $P^2=m_n^2$; in other words, those points are the first order zero of the free propagator
\footnote{
Indeed, $ c_1(n) = \alpha^\prime +2\alpha_0^\prime\ln q \frac{\cosh[(n+\beta_0-\alpha_0^\prime m_n^2+\frac{1}{2})\ln q}{\sinh(\frac{1}{2}\ln q)} \neq 0 $ ~~ for $\alpha',\alpha_0'>0 $.
}
. The wave equation Eq.(\ref{q-wave-eq}), thus, solved generally in the following form:

\begin{figure}
 \centerline{\includegraphics[width=5cm,height=3cm]{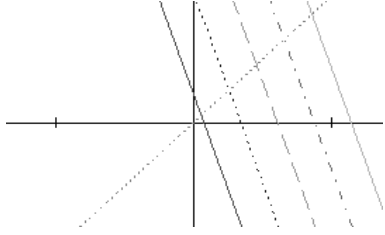}}
  \caption{The line $y=x$ vs. $y=\frac{1}{2\alpha^\prime}\{A^\dagger,A \}(x)-\frac{\omega}{\alpha^\prime}$ with $n=0,1,2,\cdots$; the mass-square eigenvalues are given as x-coordinates of the intersections.}
 \label{fig:1}
\end{figure}

\begin{equation}
 |\Phi \rangle =\sum_{n=0}^\infty \int\frac{d^3 p}{\sqrt{(2\pi)^2 2p_n^0}}\left(e^{-ip_n\cdot x}|\Phi_n\rangle c_n + h.c. \right),
\end{equation}
where $N|\Phi_n\rangle=n|\Phi_n\rangle,~P\cdot a|\Phi_n\rangle=0$ and $p_n=(\sqrt{{\mbf p}^2+m_n^2},{\mbf p}),~(n=0,1,2,\cdots)$. It should also be stressed that the wave equation does not allow
\footnote{
The eigenvalues of $P^2$ are obtained as the solutions of the equation $z=A\sinh(B-Cz)+\omega$, where $z=\alpha'P^2,A=2/\sinh(\frac{1}{2}\log q),B=(N+\beta_0+\frac{1}{2})\log q$, and $C=\omega(\alpha_0'/\alpha')\log q$. Writing $z=x+iy$ for a complex $z$, the equation can be decomposed into
\[ x=A\sinh(B-Cx)\cos y+\omega~~~{\rm and}~~~y=-A\cosh(B-Cx)\sin y ~.\]
The second equation is satisfied only for $y=0$ provided $A>0$; thus, there are no complex eigenvalues for $P^2$. 
}
 of a complex $P^2$ in addition to a space-like $P^2$.

This means that the free bi-local field is similar to local free fields in such a sense that its spacetime development can be determined by the Cauchy data without confliction with the causality. On the other hand, the Feynman propagator
\begin{equation}
 G(P^2,N)=\left(P^2+\frac{1}{2\alpha^\prime}\{A,A^\dagger\}(P^2,N)-\frac{\omega}{\alpha^\prime}+i\epsilon \right)^{-1} \label{propagator-1}
\end{equation}
decreases exponentially according as $N,|P^2| \rightarrow \infty$. As will be shown in the next section, this enables us to get a finite vacuum loop amplitude in contrast to local field theories.

\section{Interaction of the bi-local field}

\begin{figure}[t]
\hspace{10mm}
\begin{minipage}{5.5cm}
 \includegraphics[width=5cm,height=4cm]{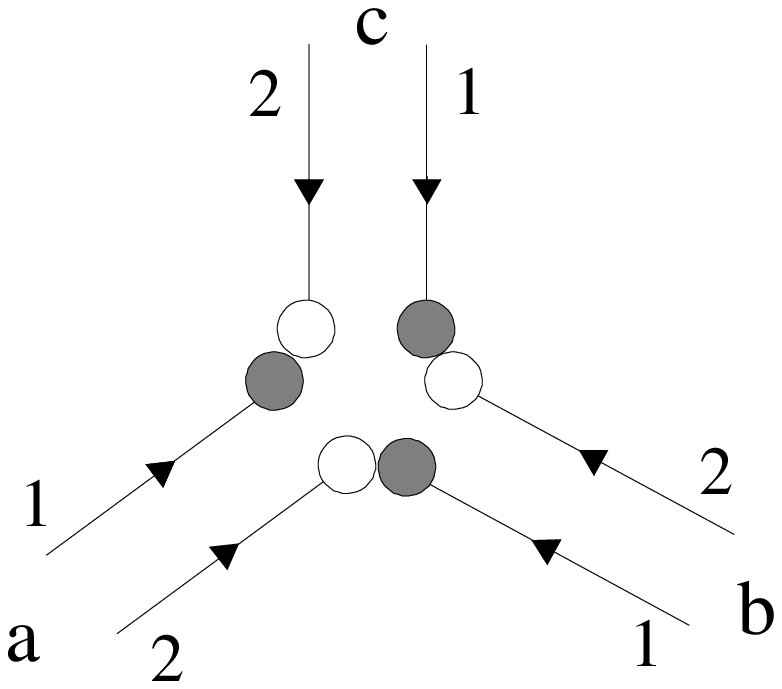}
  \caption{The a,b,c designate three bi-local systems; and 1,2 are constituents in each bi-local system.}  
  \label{fig:2}
\end{minipage}
\hspace{8mm}
\begin{minipage}{5.5cm}
 \includegraphics[width=5cm,height=4cm]{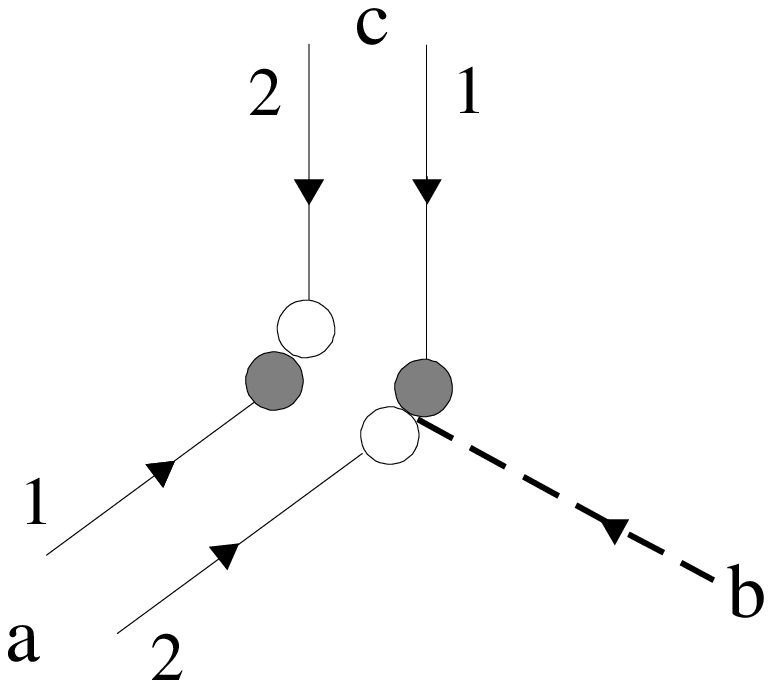}
  \caption{The dashed line denotes the local scalar field corresponding to the ground state of $b$.}
  \label{fig:3}
\end{minipage}
\end{figure}

As discussed in the previous section, we understand the q-deformation as a way getting a new dynamical system by the mapping $(a_\mu, a_\mu^\dagger)\rightarrow (A_\mu, A_\mu^\dagger)$. In other words, the physical meaning of the variables $X^\mu$ and $\bar{x}^\mu$ is not changed, but a modification is down only for the wave equation of free bi-local field. Then, the vertex function $|V \rangle$ corresponding to Fig.\ref{fig:2} should be determined by the conditions\cite{Goto-Naka}

\begin{eqnarray}
~[ x^{(1)}(b) - x^{(2)}(a) ] |V \rangle & = & 0 \nonumber \\ 
~[ p^{(1)}(b) + p^{(2)}(a) ] |V \rangle & = & 0 \label{3-vertex-1}\\
 (a,b,c~~{\rm cyclic})~~~~~~ && \nonumber
\end{eqnarray}

\noindent
with the subsidiary conditions $P(i)\cdot a(i)|V\rangle =0,(i=a,b,c)$. 

The explicit form of the vertex operator can be obtained by a standard way. In what follows, however, for the sake of simplicity, we consider a simpler case such that the particle \lq b\rq ~in  Fig.\ref{fig:3} is the ground state of the system, which can be identified with an external scalar field. Then the vertex conditions (\ref{3-vertex-1}) become

\begin{eqnarray}
~[ x^{(1)}(a) - x^{(2)}(c) ] |V \rangle & = & [ x^{(2)}(a) - x^{(1)}(c) ] |V \rangle = 0 \nonumber \\ 
~[ p^{(1)}(a) + p^{(2)}(c) ] |V \rangle & = & [ p^{(2)}(a) + p^{(1)}(c)+p(b) ] |V \rangle = 0 , \label{3-vertex-2}
\end{eqnarray}
where $p(b)$ is the momentum of the external scalar field. The conditions (\ref{3-vertex-2}) with $P(i)\cdot a(i)|V\rangle =0,(i=a,c)$ can be solved easily; and, in $p$-representation for center of mass variables, we obtain 
\begin{eqnarray}
 |V\rangle = &g& \delta^{(4)}(P(a)+P(b)+P(c)) \\ \nonumber
 & \times & e^{-\frac{i}{2\sqrt{2\kappa}}(a(a)_\perp^\dagger-a(c)_\perp^\dagger)\cdot P(b)+a(a)_\perp^\dagger\cdot a(c)_\perp^\dagger}|0\rangle , \label{vertex-1}
\end{eqnarray}
where $g$ is the coupling constant. The $|0\rangle=|0_a\rangle\otimes |0_c\rangle$ is the product of the ground states defined by $a(i)_\mu|0_i\rangle=0_i$ and $\langle 0_i|0_i \rangle =1~(i=a,c)$. Further, the projection of $a(i)_\mu$ to its physical components is written as $a_\perp(i)_\mu=O_{\mu\nu}(i)a^\nu(i)$ with $O_{\mu\nu}(i)=\eta_{\mu\nu}-P_\mu(i)P_\nu(i)/P^2(i)~(i=a,c)$. If we remark that  
\begin{eqnarray}
 && \langle \phi_a\{a(a)\}|\otimes \langle \phi_c\{a(c)\} | e^{-\frac{i}{2\sqrt{2\kappa}}(a(a)_\perp^\dagger-a(c)_\perp^\dagger)\cdot P(b)+a(a)_\perp^\dagger\cdot a(c)_\perp^\dagger}|0\rangle \nonumber \\
 &=& \langle \phi_c\{a(c)\}|e^{\frac{i}{2\sqrt{2\kappa}}a(c)_\perp^\dagger\cdot P(b)}
 \langle 0_a|e^{a(a)_\perp\cdot a(c)_\perp^\dagger}|0_c \rangle e^{\frac{i}{2\sqrt{2\kappa}}a(a)_\perp\cdot P(b)}|\phi_a\{a(a)^\dagger\}\rangle \nonumber \\
 &=& \langle \phi_c\{a(c)\}|:e^{\frac{i}{2\sqrt{2\kappa}}(a(c)^\dagger+a(c))_\perp\cdot P(b)}:|\phi_a\{a(c)^\dagger\} \rangle ,
 \end{eqnarray}
we obtain another expression of the vertex operator
\begin{equation}
 \tilde{V}(c,b,a)=g\delta^{(4)}(P(a)+P(b)+P(c)):e^{\frac{i}{2}\bar{x}^{(1)}(c)_\perp\cdot P(b)}: , \label{vertex-2}
\end{equation}
in which the operators $(a(a),a(a)^\dagger)$ are identified as $(a(c),a(c)^\dagger)$. In this case, we may confine our discussion to the physical space constructed out of $a_\perp^\dagger$ and $|0\rangle$; and then, we can write the propagator (\ref{propagator-1}) in the following form:
\begin{equation}
 G(P^2,N_\perp) = \int_C \frac{d\zeta}{2\pi i\zeta} \sum_{n=0}^\infty \zeta^{(N_\perp - n)} G(P^2,n), \label{propagator-2}
\end{equation}
where $N_\perp=-a_\perp^\dagger\cdot a_\perp$. The integral of the complex variable $\zeta$ is taken along a closed contour $C$ surrounding $\zeta=0$; for the later use, we set $C$ as a circle such as $|\zeta|<1$.

\begin{figure}
 \centerline{\includegraphics[width=4.5cm,height=5cm]{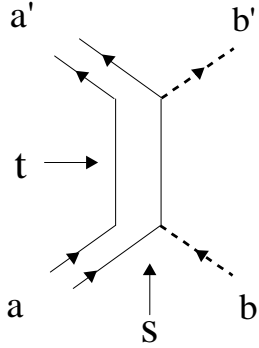}}
  \caption{A second order scattering amplitude of bi-local system by two external fields with $s=(P(a)+P(b))^2$ and $t=(P(a)-P(a'))^2$.}
 \label{fig:4}
\end{figure}

Using the above propagator and the vertex operator (\ref{vertex-1}) or (\ref{vertex-2}), first, let us  calculate the scattering amplitude corresponding to Fig.{\ref{fig:4}}. For convenience, we put $a$ and $a'$ in Fig.\ref{fig:4} as the coherent states $|z_a\rangle=e^{-a(a)_\perp^\dagger\cdot z_a}|0\rangle$ and $|z_{a'}\rangle=e^{-a(a')_\perp^\dagger\cdot z_{a'}}|0\rangle$, respectively. 

\begin{figure}[t]
\hspace{10mm}
\begin{minipage}{5cm}
 \includegraphics[width=4cm,height=4cm]{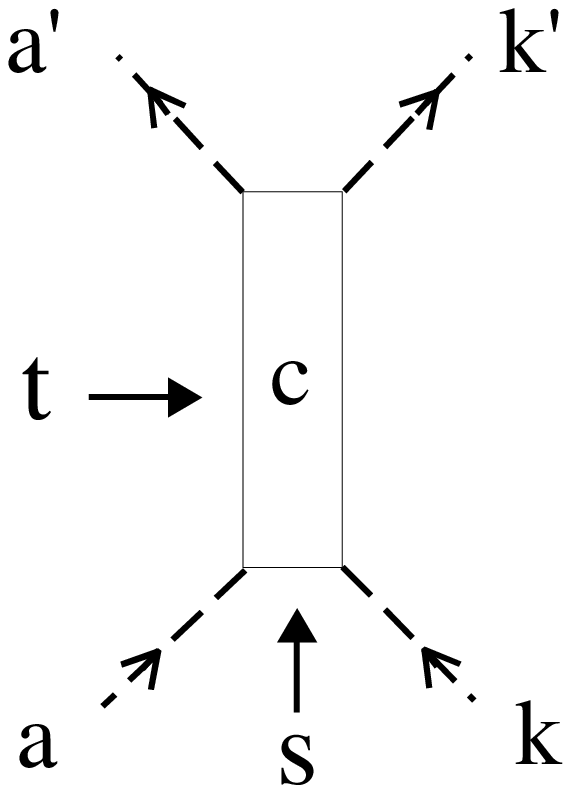}
  \caption{The s-channel scattering amplitude by four ground states.}  
  \label{fig:5}
\end{minipage}
\hspace{8mm}
\begin{minipage}{5cm}
 \includegraphics[width=5cm,height=4cm]{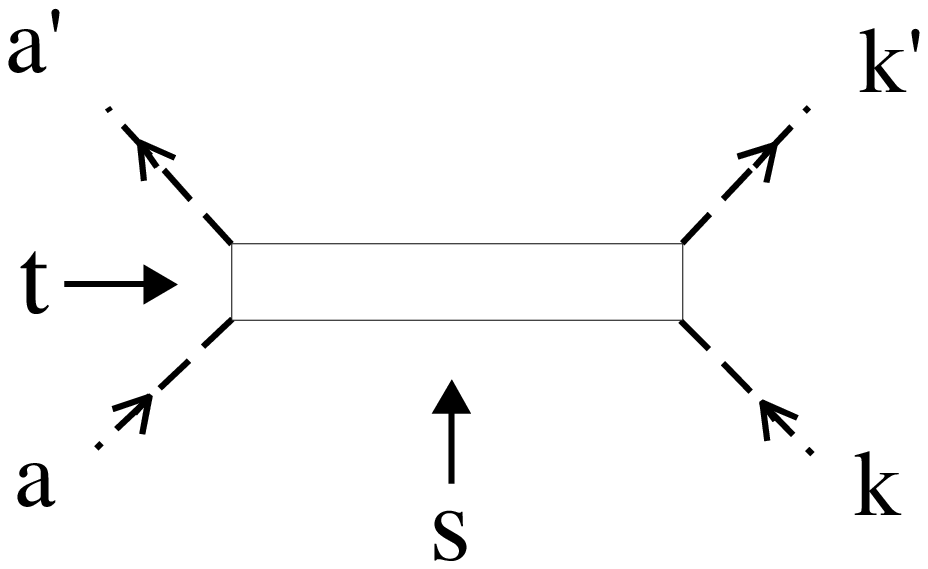}
  \caption{The t-channel amplitude obtained by the interchange $s \leftrightarrow t$ from Fig.\ref{fig:6}.}
  \label{fig:6}
\end{minipage}
\end{figure}

Then, we have the second order scattering amplitude 
\begin{eqnarray}
 A^{(2)} &=& \langle z_{a'}|\tilde{V}(a',b',c)G(s,N(c)_\perp)\tilde{V}(c,b,a)|z_a\rangle \nonumber \\
 &=& g^2\delta^4(P(a')+P(b')-P(a)-P(b))A^{(2)}_s, \label{scattering-1}
\end{eqnarray}
where
\begin{eqnarray} 
 A^{(2)}_s &=& e^{-\frac{i}{2\sqrt{2\kappa}}(z_{\perp a'}^*\cdot P(b')-z_{\perp a}\cdot P(b))}\sum_{n=0}^\infty G(s,n)\times \nonumber \\
 & \times & \frac{[(z_{a'}^*+\frac{i}{2\sqrt{2\kappa}}P(a'))\cdot O(c)\cdot (z_a-\frac{i}{2\sqrt{2\kappa}}P(a))]^n}{n!}. \label{scattering-2}
\end{eqnarray}
The summation with respect to $n$ in Eq.(\ref{scattering-2}) is convergent for a fixed $s$, since $G(s,k)\sim -\sinh(\frac{1}{2}\log q)e^{-(n+\beta_0-\alpha'_0 s+\frac{1}{2})\log q}$ for $(n+\beta_0-\alpha'_0 s+\frac{1}{2})\log q \gg 1$. In particular, for the simpler case $z_a=z_{a'}=0$ corresponding to Fig.\ref{fig:5}, the amplitude takes a simple form
\begin{equation}
 A^{(2)}_s =\sum_{n=0}^\infty G(s,n)\frac{[\frac{1}{8\kappa}(\frac{2m_0^2-t)}{2}-\frac{s}{2})]^n}{n!}. \label{scattering-3}
\end{equation}
Then, one can see that in the limit $s \rightarrow \infty$ while $t$ fixed,  the $n$-th term in the right-hand side decreases rapidly as $\sim (-\frac{\alpha'}{2}s)^ne^{-\alpha'_0 s\log q}$, since $G(s,n)\sim q^n\sinh(\frac{1}{2}\log q)e^{-(\alpha'_0 s-\beta_0-\frac{1}{2})\log q}$ for $\alpha'_0 s\log q \gg 1$. On the other hand, in the limit $t \rightarrow \infty$ while $s$ fixed, we can write
\begin{equation}
 A^{(2)}_s \sim \alpha'\sum_{n=0}^\infty \frac{1}{i}\int_0^\infty d\tau e^{i\tau \alpha'G(s,n)^{-1}} \frac{1}{n!}\left(-\frac{1}{2}\alpha' t\right)^n . \label{scattering-3}
\end{equation}
I should be notice that $G(s,n)$ decreases rapidly according as $|n-\alpha_0(s)|$ increase, where $\alpha_0(s)=\alpha'_0 s-\beta_0-\frac{1}{2}$. Hence, if we approximate $\alpha'G(s,n)^{-1} \simeq \alpha(s)-\dot{\alpha}(n-\alpha_0(s))$ with $\alpha(s)=\alpha' s-\omega,~\alpha_0(s)=\alpha'_0 s-\beta_0 -\frac{1}{2}$ and $\dot{\alpha}=1/\sinh(\frac{1}{2}\log q)$, then summation with respect to $k$ in Eq.(\ref{scattering-3}) will give the factor $\exp(-\frac{1}{2}\alpha' t e^{-i\tau\dot{\alpha}})$. Then, evaluating the integral with respect to $\tau$ by the method of steepest descent, we obtain
\begin{equation}
 A^{(2)}_s \sim F(s)\times (\alpha^\prime t)^{\alpha_1(s)} ,
\end{equation}
where $F(s)$ is a function of $s$, and $\alpha_1(s)=(\frac{\alpha^\prime}{\dot{\alpha}}+\alpha^\prime_0)s-(\frac{\omega}{\dot{\alpha}}+\beta_0+\frac{1}{2})$. This means that the $t$-channel amplitude $A^{(2)}_t$ in Fig.\ref{fig:6} obtained by interchanging $s$ and $t$ from $A^{(2)}_s$ shows the Regge behavior for a large $s$ while $t$ fixed.

\begin{figure}
 \centerline{\includegraphics[width=5cm,height=3cm]{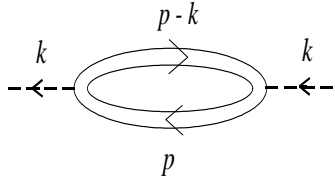}}
  \caption{The self-nenergy diagram for the ground state.}
 \label{fig:7}
\end{figure}

As the final of this section, let us consider the loop diagram Fig.\ref{fig:7} corresponding to the self-energy $\delta m^2$ for the ground state, which can be written as

\begin{align}
 \delta m^2 &\sim g^2\int d^4p Tr\left[G((p-k)^2,N_\perp(p-k)):e^{-\frac{i}{2}\bar{x}\cdot O(p-k)\cdot k}:G(p^2,N_\perp(p)):e^{\frac{i}{2}\bar{x}\cdot O(p)\cdot k}: \right] \nonumber \\
 &= \delta m^2(|p|\lnsim |k|) + \delta m^2(|p|\gg |k|) . \label{self-energy}
\end{align}
The first term of the r.h.s in Eq.(\ref{self-energy}) will be finite; and the second term can be roughly evaluated as

\begin{align}
 \delta m^2(|p| \gg |k|) &\sim g^2\int d^4p \int_C \frac{d\zeta}{2\pi i\zeta} \int_C \frac{d\zeta'}{2\pi i\zeta'} \sum_{n=0}^\infty \sum_{n'=0}^\infty \times \nonumber \\
 &Tr_{phys} \left[\zeta^{N_\perp(p)}:e^{-\frac{i}{2}\bar{x}\cdot O(p)\cdot k}:\zeta'^{N_\perp(p)}:e^{\frac{i}{2}\bar{x}\cdot O(p)\cdot k}: \right]\zeta^{-n}G(p^2,n)\zeta'^{-n'}G(p^2,n') , \label{self-energy-2}
\end{align}
where $Tr_{phys}[\cdots]$ means the \lq trace\rq in the physical subspace defined by Eq.(\ref{subsidiary}). Since the operators in Eq.(\ref{self-energy-2}) are constructed out of $(a_\perp,a_\perp^\dagger)$, the trace can be calculated by using the coherent state\cite{Coherent} $|z\rangle =e^{-z\cdot a^\dagger}|0\rangle$ as follows

\begin{align}
 Tr_{phys}[\cdots] &= \int\left(\prod_{\mu=0}^3\frac{d^2z^\mu}{\pi}\right) e^{\bar{z}^*\cdot z}\langle \bar{z}|\cdots|z\rangle |_{z_\parallel=0} \nonumber \\
 &= \frac{1}{(1-\zeta\zeta')^4}\exp \left[ -\frac{m_0^2}{2p^2}\frac{2\zeta\zeta'-(\zeta + \zeta')}{1-\zeta\zeta'}\right] \simeq \frac{1}{(1-\zeta\zeta')^4}, \label{trace}
\end{align} 
where $\bar{z}=(-z^0,z^1,z^2,z^3)$ and $z_\parallel=\frac{P(P\cdot z)}{P^2}$. As a function of $\zeta (\zeta')$, the last form of Eq.(\ref{trace}) is singular at $\zeta=\zeta'^{-1} (\zeta'=\zeta^{-1})$, which is, however, located in outside of the counter $C$. 

Further, since $G(p^2,n)$ has no pole of imaginary $p^0$, we can evaluate the integral with respect to $p$ in Eq.(\ref{self-energy-2}) by means of analytic continuation $p=(i\bar{p}^0,\bar{p}^i)$. Thus, approximating $G(-\bar{p}^2,n) \simeq -\alpha'\sinh(\frac{1}{2}\log q)e^{-(n+\beta_0+\alpha_0'\bar{p}^2+\frac{1}{2})\log q}$ for large $\bar{p}$, the summation with respect to $n,n'$ gives rise to $[\alpha'\sinh(\frac{1}{2}\log q)e^{-(\beta_0+\alpha_0'\bar{p}^2+\frac{1}{2})\log q}]^2(1-\zeta q^{-1})(1-\zeta'q^{-1})$. Then we can carried out the $p$-integral in Eq.(\ref{self-energy-2}) so that
\begin{align}
 \delta m^2(|p| \gg |k|) & \sim i \left(\frac{g\pi\alpha'\sinh(\frac{1}{2}\log q)}{2\alpha_0'\log q}e^{-(\beta_0+\frac{1}{2})\log q}\right)^2 \nonumber \\
 & \int_C\frac{d\zeta}{2\pi i\zeta} \int_C\frac{d\zeta'}{2\pi i\zeta'}\frac{1}{(1-\zeta q^{-1})(1-\zeta' q^{-1})(1-\zeta\zeta')^4}. \label{self-energy-3} 
\end{align}
The integral with respect to $\zeta$ and $\zeta'$ comes to be $1$ provided that the radius of the contour $C$ is smaller than $q$; therefore, the self energy in our model is be convergent
\footnote{
If we do not ignore $m_0^2/p^2$ term in the exponential in Eq.(\ref{trace}), then the substitution
\[ 1\Big/(1-\zeta q^{-1})(1-\zeta' q^{-1})(1-\zeta\zeta')^4 \rightarrow
 \exp \left[ -8\sqrt{-(\alpha_0'm_0^2\log q)\frac{2\zeta\zeta'-(\zeta + \zeta')}{1-\zeta\zeta'}}\right] \Big/ (1-\zeta q^{-1})(1-\zeta' q^{-1})(1-\zeta\zeta')^4 \]
must be done. In this case, the integral in Eq.(\ref{self-energy-3}) is still convergent.}
 .

\section{Summary and discussions}

The relativistic two-particle system, the bi-local system, bounded by 4-dimensional harmonic oscillator potential yields a successful description of two-body meson like states. The mass square spectrum, then, arises from the excitation of relative variables, which are independent of center of mass variables. 

In this paper, we have tried to construct a q-deformed bi-local system in such a way that the center of mass momenta are included in the deformation parameters of relative variables. In other words, the q-deformed relative variables become functions of the center of mass momenta and of original relative variables, which are independent of center of mass variables. Then, the q-deformed relative coordinates become non-commutative, while the center of mass coordinates remain as commutative variables. As a result of this deformation, the mass square operator gives rise to a non-linear mass square spectrum. The way of deformation is not unique, and we have defined it from a heuristic point of view such that the mass square operator becomes a Lorentz scalar in spite of non-covariant property of oscillator variables.

In the q-deformed bi-local system, the wave function of the system acquires new aspects such that the propagator of the system dumps rapidly according as $|P^2|$ or $N$ tends to infinity. With that in mind, further, we have studied the interaction of the bi-local system with external scalar fields, which are identified with the ground state of the system. To this end, we have defined the three vertexes among two bi-local systems and one scalar field; and then, we have verified the flowing: First, the second order t-channel scattering amplitude shows the Regge behavior in the limit $t \rightarrow \infty$ while $s$ fixed. Secondly, we have calculated a second-order self-energy diagram, and it is shown that the self energy diagram of the bi-local field comes to be convergent due to the characteristic property of the propagator. 

The convergence problem, however, is still subtle, since it is necessary to choose a suitable contour in the integral representation of the propagator. In order to fix the field theoretical properties in the q-deformed bi-local system, it will be necessary to study the higher order diagrams. Further, for interacting case, the problem of causality is remained as an open question, since the field equation contains higher order derivatives with respect to time parameter. In addition to those, the way of q-deformation (\ref{mapping}) for 4-dimensional oscillator variables is not sufficient because of its non covariant property. Thus, it should also be required to study the guiding principle defining the deformation from various points of view. These are interesting and important subjects for a future study.

\section*{Acknowledgements}

The authors wish to express their thanks to the members of their laboratory for discussions and encouragement.

\appendix

\section{Representation of a q-oscillator}

We here review the representation of the q-oscillator variables defined by
\begin{equation}
 [A,A^\dagger]_q \equiv A A^\dagger -q^\alpha A^\dagger A =q^{-\alpha ( N + \beta) }, \label{q-commutator}
\end{equation}
where $\alpha,\beta$ and $q (\geq 1)$ are real parameters. The $N=a^\dagger a$ is the number operator for the ordinary oscillator variables satisfying $[a,a^\dagger]=1$. The explicit mapping between $(a,a^\dagger)$ and $(A,A^\dagger)$ is given by
\begin{equation}
 A = a \sqrt{\frac{[N]_q}{N}},~~A^\dagger = \sqrt{\frac{[N]_q}{N}}a^\dagger ,
\end{equation}
where $N=a^\dagger a$, and $[N]_q$ is a function of $N$ determined in what follows. Now, remembering $Na=a(N-1)$ and $Na^\dagger=a^\dagger(N+1)$, one can verify that
\begin{equation}
 A A^\dagger = [N+1]_q,~~A^\dagger A = [N]_q . 
\end{equation}
Then Eq.(\ref{q-commutator}) can be reduced to
\begin{equation}
 [N+1]_q - q^\alpha [N]_q = q^{-\alpha (N +\beta) } .  \label{recurrence}
\end{equation}
The recurrence equation (\ref{recurrence}) can be solved easily in the following form:
\begin{equation}
 [N]_q = q^{-\alpha\beta } \frac{q^{\alpha N} - q^{-\alpha N}}{q^\alpha - q^{-\alpha}} + q^{\alpha N}[0]_q , 
\end{equation}
where $[0]_q$ is an arbitrary initial term for $N=0$. We, here, choose
\begin{equation}
 [0]_q= \frac{q^{\alpha\beta}-q^{-\alpha\beta}}{q^\alpha - q^{-\alpha}}
\end{equation}
; then, $[N]_q$ has the simple form
\begin{equation}
 [N]_q = \frac{q^{\alpha(N+\beta)}-q^{-\alpha(N+\beta)}}{q^\alpha -q^{-\alpha}} \label{[N]}
\end{equation}
After this q-deformation, the hamiltonian type of combination $\frac{1}{2}\{ a^\dagger ,a \}+\beta$ of the ordinary oscillators should be replaced with $\frac{1}{2} \{ A^\dagger ,A \}$, to which we have the expression
\begin{eqnarray}
 \frac{1}{2} \{ A^\dagger,A \} &=& \frac{1}{2}\left( [N]_q + [N+1]_q \right) \nonumber \\
  &=& \frac{1}{2}\frac{\sinh \left[\alpha(N + \beta +\frac{1}{2})\log q \right]}{\sinh(\frac{1}{2}\alpha\log q)}. \label{q-hamiltonian}
\end{eqnarray}
Indeed, one can verify that the $\frac{1}{2} \{ A,A^\dagger \}$ becomes $\frac{1}{2}\{ a^\dagger ,a \}+\beta$ according as $q \rightarrow 1$. This implies that the $\beta$ plays the role of an additional factor to the zero point energy of the four-dimensional oscillator for $q=1$; on the other hand, the $\alpha$ can be absorbed in the definition of $q$ by the substitution $q^\alpha \rightarrow q$. 

The q-deformation can be extended to N-dimensional oscillator variables, though there arise several problems. Let us consider the D-dimensional oscillator variables defined by $[a_i,a_j^\dagger]=\delta_{ij},[a_i,a_j]=[a\i^\dagger,a_j^\dagger]=0~(i,j=1,2,\cdots,D)$. We have to remarked that there is no mapping $A_i(a,a^\dagger)$ satisfying
\begin{equation}
 [A_i,A_j^\dagger]_q=\delta_{ij}f(N)~~{\rm and}~~[A_i,A_j]_q=[A_i^\dagger,A_j^\dagger]_q=0,
\end{equation}
where $N=\sum_ia_i^\dagger a_i$. Indeed, since $A_j[A_i,A_j^\dagger]_q=q^{2\alpha}[A_i,A_j^\dagger]_q A_j+q^\alpha [A_i,f(N)]$ for $i\neq j$, we have $[A_i,f(N)]=0$, which holds only when $A_i$ is a function of $N$; that is, a function without vector indices $\{i\}$.  The followings, however, may be available mappings: \\

Case~(i)  $A_i=a_i\sqrt{\frac{[N_i]_q}{N_i}}$ \vspace{3mm} \\
In this case, we have the simple q-deformed algebra
\begin{equation}
 [A_i,A_j^\dagger]_q =\delta_{ij}q^{-\alpha ( N_i + \beta) }, \label{q-commutator-2}
\end{equation}
which spoils, however, $U(D)$ symmetry even for $\sum_i\{A_i,A_i^\dagger\}$.

Case~(ii)  $A_i=a_i\sqrt{\frac{[N]_q}{N}}$ \vspace{3mm} \\
This is a covariant mapping which keeps the $U(D)$ vector property of $A_i$. The algebra is, however, not simple so that 
\begin{equation}
 [A_i,A_j^\dagger]_q=\frac{[N+1]_q}{N+1}\delta_{ij}+\left(\frac{[N+1]_q}{[N]_q}\frac{N}{N+1}-q^\alpha \right)A_j^\dagger A_i
\end{equation}
In this case, the hamiltonian type of combination becomes
\begin{equation}
 \frac{1}{2}\sum_i\{ A_i^\dagger,A_i \} = \frac{1}{2}\frac{\sinh \left[\alpha(N + \beta +\frac{1}{2})\log q \right]}{\sinh(\frac{1}{2}\alpha\log q)}+\frac{1}{2}\frac{D+1}{N+1}\frac{\sinh \left[\alpha(N + \beta +1)\log q \right]}{\sinh(\alpha\log q)}.
\end{equation}

Case~(iii) $A_i = a_i\sqrt{\frac{[N]_q}{N_i}}$ \vspace{3mm} \\
The mapping looks like to break $U(D)$ symmetry of the q-deformed algebra, but we have
\begin{equation}
 [A_i,A_j^\dagger]_q = q^{-\alpha(N+\beta)}A_i[N]_q^{-1}A_j^\dagger \label{commutator-3}
\end{equation}
and the simple hamiltonian type of combination
\begin{equation}
\frac{1}{2}\sum_i\{ A_i^\dagger,A_i \}=\frac{D}{2}\frac{\sinh \left[\alpha(N + \beta +\frac{1}{2})\log q \right]}{\sinh(\frac{1}{2}\alpha\log q)}. \label{q-hamiltonian-2}
\end{equation}
The Eq.(\ref{q-hamiltonian-2}) is the direct extension of Eq.(\ref{q-hamiltonian}), though $\{A_i\}$ do not carry D-dimensional vector property from $\{a_i\}$.

{}

\end{document}